\documentclass[russian]{jetpl}

\twocolumn

\usepackage{amsmath,amssymb,color,pstricks,bm,alltt}

\usepackage{graphicx}

\begin{document}

%%% article in English
\lat

%%% article title
\title{Antiadiabatic phonons, Coulomb pseudopotential and
superconductivity in Eliashberg -- McMillan theory}

%%% article title - for colontitle (at the top of the page)
\rtitle{Antiadiabatic phonons, Coulomb pseudopotential and
superconductivity}

%%% article title - for table of contents (usually identical with \title)
\sodtitle{Antiadiabatic phonons, Coulomb pseudopotential and
superconductivity in Eliashberg -- McMillan theory}

%%% author(s) ( + e-mail)
\author{M.\ V.\ Sadovskii\thanks{E-mail: sadovski@iep.uran.ru},
}

%%% author(s) - for colontitle (at the top of the page)
\rauthor{M.\ V.\ Sadovskii}

%%% author(s) - for table of contents
\sodauthor{M.\ V.\ Sadovskii}

%%% author's address(es)
\address{Institute for Electrophysics, RAS Ural Branch, Amundsen str. 106, 
Ekaterinburg 620016, Russia
}

%%% dates of submition & resubmition (if submitted once, second argument is *)
%\dates{}{*}

\abstract{
The influence of antiadiabatic phonons on the temperature of superconducting 
transition is considered within Eliashberg -- McMillan approach in the model 
of discrete set of (optical) phonon frequencies.
A general expression for superconducting transition temperature $T_c$ is
proposed, which is valid in situation, when one (or several) of such phonons
becomes antiadiabatic. We study the contribution of such phonons into
the Coulomb pseudopotential $\mu^{\star}$. It is shown, that antiadiabatic
phonons do not contribute to Tolmachev's logarithm and its value is
determined by partial contributions from adiabatic phonons only. The results
obtained are discussed in the context of the problem of unusually high
superconducting transition temperature of FeSe monolayer on STO.
}


\maketitle

%\newpage

\section{Introduction}

The most developed approach to description of superconductivity in the system
of electrons and phonons is Eliashberg -- McMillan theory \cite{Scal,Izy,All,Grk-Krs}.
It is well known, that this theory is completely based on the applicability of
adiabatic approximation and Migdal theorem \cite{Mig}, which allows to neglect
vertex corrections in calculations of the effects of electron -- phonon
interaction in typical metals. The real small parameter of perturbation theory
is $\lambda\frac{\Omega_0}{E_F}\ll 1$, where  $\lambda$ is the dimensionless
coupling constant of electron -- phonon interaction, $\Omega_0$ is characteristic
phonon frequency and  $E_F$ is Fermi energy of the electrons. In particular,
this leads to a conclusion, that vertex corrections in this theory can be neglected
even for $\lambda > 1$, because of the validity of inequality  
$\frac{\Omega_0}{E_F}\ll 1$ characteristic for typical metals.

In a recent paper \cite{Sad_18} we have shown, that under the conditions of 
strong nonadiabaticity , when $\Omega_0\gg E_F$, a new small parameter appears
in the theory $\lambda_D\sim \lambda\frac{E_F}{\Omega_0}\sim\lambda\frac{D}
{\Omega_0}\ll 1$ ($D$ is the halfwidth of electron band), so that corrections
to electronic spectrum become irrelevant and vertex correction can be
similarly neglected \cite{Ikeda}. In general case, the renormalization of
electronic spectrum (effective mass of an electron) is determined by the new
dimensionless constant $\tilde\lambda$, which reduces to the usual $\lambda$ 
in adiabatic limit, while in the strong antiadiabatic limit it tends to
$\lambda_D$. At the same time, the temperature of superconducting transition
$T_c$ in antiadiabatic limit is determined by Eliashberg -- McMillan
pairing coupling constant $\lambda$, while the preexponential factor in the
expression for $T_c$, which is of the typical weak -- coupling form, is
determined by band halfwidth (Fermi energy). For the case of the interaction
with a single optical phonon in Ref. \cite{Sad_18} we obtained the unified
expression for $T_c$, valid both in adiabatic and antiadiabatic regimes, and
producing a smooth interpolation in the intermediate region.

In Ref. \cite{Sad_18} we also noted, that the presence of high phonon frequencies
of the order of or even exceeding the Fermi energy, leads to the obvious
suppression of Tolmachev's logarithm in the expression for Coulomb pseudopotential
$\mu^{\star}$, which creates additional difficulties for the realization of
superconducting state in the system with antiadiabatic phonons.

The interest to this problem is stimulated by the discovery of a number 
superconductors, where adiabatic approximation is not valid, while characteristic
phonon frequencies are of the order  of or even higher than Fermi energy of
electrons. Most typical in this sense are intercalated systems with monolayers
of FeSe, as well as monolayers of FeSe on Sr(Ba)TiO$_3$ (and similar) substrates 
(FeSe/STO) \cite{UFN}. For the first time, the nonadiabatic character of
superconductivity in FeSe/STO was noted by Gor'kov  \cite{Gork_1,Gork_2}, while
discussing the idea of possible mechanism of the enhancement of superconducting
transition temperature $T_c$ in FeSe/STO system due to interaction with
high energy  optical phonons of SrTiO$_3$ \cite{UFN}.

In the present paper we consider the generalized model with discrete set of the
frequencies of (optical) phonons, part of which may be andiabatic.
We obtain the general expressions for $T_c$, valid both in adiabatic and
antiadiabatic limits. We also present the general analysis of the problem of the
Coulomb pseudopotential in such model. The results obtained are used for simple
estimates of $T_c$ in situation typical for FeSe/STO.

\section{Temperature of superconducting transition}

Linearized Eliashberg equations, determining superconducting transition
temperature $T_c$, written in real frequencies representation, have the
following form \cite{Izy}:
\begin{eqnarray}
[1-Z(\varepsilon)]\varepsilon =\int_{0}^{D}d\varepsilon'\int_{0}^{\infty}d\omega
\alpha^2(\omega)F(\omega)f(-\varepsilon')\times\nonumber\\
\times\left(\frac{1}{\varepsilon'+\varepsilon+\omega+i\delta}-
\frac{1}{\varepsilon'-\varepsilon+\omega-i\delta}\right)
\label{lin_Z}
\end{eqnarray}
\begin{eqnarray}
Z(\varepsilon)\Delta(\varepsilon)=\int_{0}^{D}\frac{d\varepsilon'}{\varepsilon'}
th\frac{\varepsilon'}{2T_c}Re\Delta(\varepsilon')\times
\nonumber\\
\times\int_{0}^{\infty}d\omega
\alpha^2(\omega)F(\omega)
\times\nonumber\\
\times\left(\frac{1}{\varepsilon'+\varepsilon+\omega+i\delta}+
\frac{1}{\varepsilon'-\varepsilon+\omega-i\delta}\right)
\label{lin_D}
\end{eqnarray}
Here $\Delta(\omega)$ is the gap function of a superconductor, while $Z(\omega)$ 
is electron mass renormalization function and $f(\varepsilon)$ is Fermi distribution.
In difference with the standard approach \cite{Izy}, we have introduced the finite
integration limits, determined by the (half)bandwidth $D$. In the following we assume
the half--filled band of degenerate electrons in two dimensions, so that
$D=E_F\gg T_c$, with constant density of states. For simplicity at first we neglect 
the contribution of direct Coulomb repulsion.
In these (integral) equations $\alpha^2(\omega)$ represents Eliashberg -- McMillan
function, determining the strength electron -- phonon interaction, and
$F(\omega)$ is the phonon density of states. Eliashberg -- McMillan coupling constant
is defined as:
\begin{equation}
\lambda=2\int_{0}^{\infty}\frac{d\omega}{\omega}\alpha^2(\omega)F(\omega)
\label{lambda_Elias_Mc}
\end{equation}
The details concerning its calculation for systems with nonadiabatic phonons
were discussed in details in Ref. \cite{Sad_18}.

Situation is considerably simplified \cite{Sad_18}, if we consider these equations
in the limit of $\varepsilon\to 0$ and look for the solutions $Z(0)=Z$ and 
$\Delta(0)=\Delta$. Then from (\ref{lin_Z}) we obtain:
\begin{equation}
[1-Z]\varepsilon=
-2\varepsilon\int_{0}^{\infty}d\omega\alpha^2(\omega)F(\omega)\frac{D}{\omega
(\omega+D)}
\label{Z_sc_eq}
\end{equation}
or
\begin{equation}
Z=1+\tilde\lambda
\label{Z_sc}
\end{equation}
where constant $\tilde\lambda$ is defined as:
\begin{equation}
\tilde\lambda=2\int_{0}^{\infty}\frac{d\omega}{\omega}\alpha^2(\omega)
F(\omega)\frac{D}{\omega+D}
\label{A6}
\end{equation}
which for $D\to\infty$ reduces to the usual Eliasberg -- McMillan constant
(\ref{lambda_Elias_Mc}), while for $D$ significantly smaller than characteristic
phonon frequencies it gives the ``antiadiabatic'' coupling constant:
\begin{equation}
\lambda_D=
2D\int \frac{d\omega}{\omega^2}\alpha^2(\omega)F(\omega)
\label{derivata_b}
\end{equation}
Eq. (\ref{A6}) describes smooth transition between the limits of wide and narrow
conduction bands. Mass renormalization is, in general case, determined exclusively
by constant $\tilde\lambda$:
\begin{equation}
m^{\star}=m(1+\tilde\lambda)
\label{mass_renrm}
\end{equation}
In the limit of strong nonadiabaticity this renormalization is quite small and
determined by the limiting expression $\lambda_D$ \cite{Sad_18}.

From Eq. (\ref{lin_D}) in the limit of $\varepsilon\to 0$ and using (\ref{Z_sc}), 
we immediately obtain the following expression for $T_c$:
\begin{equation}
1+\tilde\lambda=2\int_{0}^{\infty}d\omega\alpha^2(\omega)F(\omega)
\int_{0}^{D}\frac{d\varepsilon'}{\varepsilon'(\varepsilon'+\omega)}
th\frac{\varepsilon'}{2T_c}
\label{Tc}
\end{equation}
Consider now the situation with discrete set of phonon modes
(dispesionless, Einstein phonons). In this case the phonon density of states
is written as: 
\begin{equation}
F(\omega)=\sum_{i}\delta(\omega-\omega_i)
\label{Fwi}
\end{equation}
where $\omega_i$ are discrete frequencies modeling the optical branches of
the phonon spectrum. Then from Eqs. (\ref{lambda_Elias_Mc}) and (\ref{A6}) we have:
\begin{equation}
\lambda=2\sum_i\frac{\alpha^2(\omega_i)}{\omega_i}\equiv\sum_i\lambda_i
\label{lamb_i}
\end{equation}
\begin{equation}
\tilde\lambda=2\sum_i\frac{\alpha^2(\omega_i)D}{\omega_i(\omega_i+D)}
=2\sum_i\lambda_i\frac{D}{\omega_i+D}
\equiv\sum_i\tilde\lambda_i
\label{lamb_tild}
\end{equation}
Correspondingly, in this case:
\begin{equation}
\alpha^2(\omega)F(\omega)=\sum_i\alpha^2(\omega_i)\delta(\omega-\omega_i)=
\sum_i\frac{\lambda_i}{2}\omega_i\delta(\omega-\omega_i)
\label{El-Mc-discr}
\end{equation}
The standard Eliashberg equation (in adiabatic limit) for such model were
consistently solved in Ref. \cite{KM}. For our purposes it is sufficient to
analyze only the Eq. (\ref{Tc}), which takes now the following form:
\begin{equation}
1+\tilde\lambda=2\sum_i\alpha^2(\omega_i)
\int_{0}^{D}\frac{d\varepsilon'}{\varepsilon'(\varepsilon'+\omega_i)}
th\frac{\varepsilon'}{2T_c}
\label{Tc_opt}
\end{equation}
Solving Eq. (\ref{Tc_opt}) we obtain:
\begin{equation}
T_c\sim 
\prod_i\left(\frac{D}{1+\frac{D}{\omega_i}}\right)^{\frac{\lambda_i}{\lambda}}
\exp\left(-\frac{1+\tilde\lambda}{\lambda}\right)
\label{Tc_opt_i}
\end{equation}
For the case of two optical phonons with frequencies $\omega_1$ and $\omega_2$ we
have:
\begin{equation}
T_c\sim
\left(\frac{D}{1+\frac{D}{\omega_1}}\right)^{\frac{\lambda_1}{\lambda}}
\left(\frac{D}{1+\frac{D}{\omega_2}}\right)^{\frac{\lambda_2}{\lambda}}
\exp\left(-\frac{1+\tilde\lambda}{\lambda}\right)
\label{Tc_opt_2}
\end{equation}
where $\tilde\lambda=\tilde\lambda_1+\tilde\lambda_2$ and 
$\lambda=\lambda_1+\lambda_2$. For the case of $\omega_1\ll D$ (adiabatic phonon), 
and $\omega_2\gg D$ (antiadiabatic phonon) Eq. (\ref{Tc_opt_2}) is immediately
reduced to:
\begin{equation}
T_c\sim (\omega_1)^{\frac{\lambda_1}{\lambda}}
(D)^{\frac{\lambda_2}{\lambda}}
\exp\left(-\frac{1+\tilde\lambda}{\lambda}\right)
\label{Tc_opt_ad_ant}
\end{equation}
Here we can see, that in the preexponential factor the frequency of antiadiabatic
phonon is replaced by band halfwidth (Fermi energy), which plays a role of cutoff 
for logarithmic divergence in Cooper channel in antiadiabatic limit 
\cite{Sad_18,Gork_1,Gork_2}.

The general result (\ref{Tc_opt_i}) gives the unified expression for $T_c$ for the
discrete set of optical phonons, valid both in adiabatic and antiadiabatic
regimes and interpolating between these limit in intermediate region.

\section{Coulomb pseudopotential}

Above we had neglected the direct Coulomb repulsion of electrons, which in the
standard approach \cite{Scal,Izy,All} is described by Coulomb pseudopotential
$\mu^{\star}$, which is effectively suppressed by large Tolmachev's logarithm.
As was noted in Ref. \cite{Sad_18} antiadiabatic phonons suppress Tolmachev's
logarithm, which apparently leads to a sufficient suppression of the temperature
of superconducting transition. To clarify this situation let us consider the
simplified version of integral equation for the gap (\ref{lin_D}), writing it as:
\begin{equation}
Z(\varepsilon)\Delta(\varepsilon)=
\int_{0}^{D}d\varepsilon'K(\varepsilon,\varepsilon')\frac{1}{\varepsilon'}
th\frac{\varepsilon'}{2T_c}\Delta(\varepsilon')
\label{Tcc}
\end{equation}
where the integral kernel we write as a combination of two step -- functions:
\begin{equation}
K(\varepsilon,\varepsilon')=\lambda\theta(\tilde D-|\varepsilon|)
\theta(\tilde D-|\varepsilon'|)-\mu\theta(D-|\varepsilon|)\theta(D-|\varepsilon'|)
\label{K2step}
\end{equation}
where $\mu$ is the dimensionless (repulsive) Coulomb potential, while the
parameter $\tilde D$, determining the energy width of attraction region due to
phonons is determined by preexponential factor of Eq. (\ref{Tc_opt_i}):
\begin{equation}
\tilde D=\prod_i\left(\frac{D}{1+\frac{D}{\omega_i}}\right)^{\frac{\lambda_i}
{\lambda}}
\label{tildaD}
\end{equation}
Note that we always have $\tilde D<D$. Eq. (\ref{Tcc}) is now rewritten as:
\begin{equation}
Z(\varepsilon)\Delta(\varepsilon)=(\lambda-\mu)\int_{0}^{\tilde D}
\frac{d\varepsilon'}{\varepsilon'}th\frac{\varepsilon'}{2T_c}\Delta(\varepsilon')
-\mu\int_{\tilde D}^{D}\frac{d\varepsilon'}{\varepsilon'}\Delta(\varepsilon')
\label{Tccc}
\end{equation}
Writing the mass renormalization due to phonons as:
\begin{equation}
Z(\varepsilon)=\left\{\begin{array}{c} 
1+\tilde\lambda\quad\mbox{for}\quad\varepsilon<\tilde D \\
1\quad\mbox{for}\quad\varepsilon>\tilde D
\end{array}
\right.
\label{Zmassph}
\end{equation}
we look for the solution of Eq. (\ref{Tcc}) for $\Delta(\varepsilon)$, as usual, 
also in two -- step form \cite{Scal,Izy,All}:
\begin{equation}
\Delta(\varepsilon)=\left\{\begin{array}{c} 
\Delta_1\quad\mbox{for}\quad\varepsilon<\tilde D \\
\Delta_2\quad\mbox{for}\quad\varepsilon>\tilde D
\end{array}
\right.
\label{Delta2step}
\end{equation}
Then Eq. (\ref{Tccc}) transforms into the system of two homogeneous linear
equations for constants $\Delta_1$ and $\Delta_2$:
\begin{eqnarray}
(1+\tilde\lambda)\Delta_1=(\lambda-\mu)\ln\frac{\tilde D}{T_c}\Delta_1
-\mu\ln\frac{D}{\tilde D}\Delta_2\nonumber\\
\Delta_2=-\mu\ln\frac{\tilde D}{T_c}\Delta_1-\mu\ln\frac{D}{\tilde D}\Delta_2
\label{D1D2}
\end{eqnarray}  
with the condition for nontrivial solution taking the form:
\begin{equation}
1+\tilde\lambda=\left(\lambda-\frac{\mu}{1+\mu\ln\frac{D}{\tilde D}}\right)
\ln\frac{\tilde D}{T_c}
\label{Detsys}
\end{equation}
Correspondingly, for the transition temperature we get:
\begin{equation}
T_c=\tilde D\exp\left(-\frac{1+\tilde\lambda}{\lambda-\mu^{\star}}\right)
\label{Tcccc}
\end{equation}
where the Coulomb pseudopotential is determined by the following expression:
\begin{equation}
\mu^{\star}=\frac{\mu}{1+\mu\ln\frac{D}{\tilde D}}=
\frac{\mu}{1+\mu\ln\prod_i\left(1+\frac{D}{\omega_i}\right)^{\frac{\lambda_i}{\lambda}}}
\label{mustar}
\end{equation}
Thus, the phonon frequencies enter Tolmachev's logarithm as the product of
partial contributions, with values determined also by corresponding coupling
constants. Similar structure of Tolmachev's logarithm was first obtained
(in somehow different model) in Ref. \cite{KMK}, where the case of frequencies
going outside the limits of adiabatic approximation was not considered.
In this sense, Eq. (\ref{mustar}) has a wider region of applicability.
In particular, for the model of two optical phonons with frequencies
$\omega_1\ll D$ (adiabatic phonon) and $\omega_2\gg D$, from Eq. (\ref{mustar}) 
we get:
\begin{equation}
\mu^{\star}=\frac{\mu}{1+\mu\ln\left(\frac{D}{\omega_1}
\right)^{\frac{\lambda_1}{\lambda}}}=
\frac{\mu}{1+\mu\frac{\lambda_1}{\lambda}\ln\frac{D}{\omega_1}}
\label{mstar}
\end{equation}
We can see, that the contribution of antiadiabatic phonon drops out of
Tolmachev's logarithm, while the logarithm itself remains, with its value
determined by the ratio of the band halfwidth (Fermi energy) to the frequency
of adiabatic (low frequency) phonon. The general effect of suppression of
Coulomb repulsion also remains, though it becomes weaker proportionally to
to the partial interaction of electrons with corresponding phonon.
This situation is conserved also in the general case --- the value of
Tolmachev's logarithm and corresponding Coulomb pseudopotential is
determined by contributions of adiabatic phonons, while antiadiabatic phonons 
drops out. Thus, in general case, situation becomes more favorable for
superconductivity, as compared to the case of a single antiadiabatic phonon,
considered in Ref. \cite{Sad_18}.

\section{Conclusions}

In the present paper we have considered the electron -- phonon coupling in
Eliashberg -- McMillan theory in situation, when antiadiabatic phonons with
high enough frequency (comparable or exceeding the Fermi energy $E_F$) are 
present in the system. The value of mass renormalization, in general case,
is determined by the coupling constant $\tilde\lambda$, while the value of
the pairing interaction is always determined by the standard coupling constant
$\lambda$ of Eliashberg -- McMillan theory, appropriately generalized by taking
into account the finite value of phonon frequency  \cite{Sad_18}. 
Mass renormalization due to antiadiabatic phonons is small and determined by
the coupling constant $\lambda_D\ll\lambda$. In this sense, in the limit of strong
antiadiabaticity, the coupling of such phonons with electrons becomes weak and
corresponding vertex correction are irrelevant \cite{Ikeda}, similarly to the case
of adiabatic phonons \cite{Mig}. Precisely this this fact allows us to use 
Eliashberg -- McMillan approach in the limit of strong antiadiabaticity. 
In the intermediate region all expressions proposed above are of interpolating
nature and for more deep understanding of this region we have to use other
approaches (see e.g. Refs. \cite{Gri_1,Gri_2}).

The cutoff of pairing interaction in Cooper channel in antiadiabatic limit
takes place at energies $\sim E_F$, as was previously noted in 
Refs. \cite{Sad_18,Gork_1,Gork_2}), so that corresponding phonons do not
contribute to Tolmachev's logarithm in Coulomb pseudopotential, though large
enough values of this logarithm (and corresponding smallness of $\mu^{\star}$) 
can be guaranteed due to contributions from adiabatic phonons.

Note that above we have used rather simplified analysis of Eliashberg
equations. However, in our opinion, more elaborate approach, e.g. along the
lines of Ref. \cite{KM}, will not lead to qualitative change of our results.

\begin{figure}
\includegraphics[clip=true,width=0.45\textwidth]{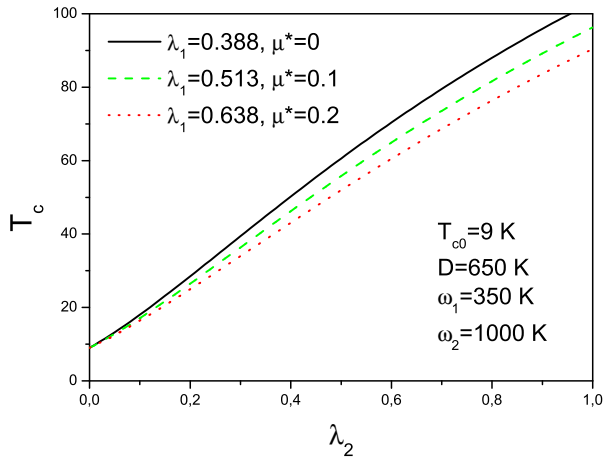}
\caption{Fig. 1. Dependence of superconducting transition temperature on the
coupling constant with high -- energy phonon for the typical values of
parameters of FeSe/STO system. 
}
\label{Tc_comp}
\end{figure}

In conclusion let us discuss the current results in the context of possible
explanation of high -- temperature superconductivity in a monolayer of FeSe
on Sr(Ba)TiO$_3$ (FeSe/STO) \cite{UFN}. The presence in Sr(Ba)TiO$_3$ of
high -- energy optical phonons indicates the possibility of significant
enhancement of $T_c$ in this system due to interactions of FeSe electrons 
with these phonons on FeSe/STO interface \cite{UFN,FeSe_ARPES_Nature}. 
ARPES experiments \cite{FeSe_ARPES_Nature} and LDA+DMFT calculations 
\cite{NPS_1,NPS_2} have shown, that Fermi energy $E_F$ in this system is
significantly (practically two times) lower than the energy of the optical
phonon, which unambigously indicates the realization, in this case, of
antiadiabatic situation \cite{Gork_1,Gork_2}. Let us look if we can explain
the observed high values of $T_c$ in this system using the expressions derived
in this work. Assuming for FeSe on STO the characteristic value of phonon
frequency $\omega_1=$ 350K, Fermi energy $E_F=D=$ 650K, and the energy of the
optical phonon in SrTiO$_3$ $\omega_2=$ 1000K \cite{UFN,FeSe_ARPES_Nature}, 
we calculate $T_c$ using Eqs. (\ref{Tc_opt_2}), (\ref{Tcccc}) (the case two
phonon frequencies), considering $\mu^{\star}$ as a free model parameter. 
Let us choose the value of $\lambda_1$ to obtain, in the absence of interactions
with high -- energy phonon of STO, the value of $T_c=$ 9K, typical for the
bulk FeSe, which gives $\lambda_1>$ 0.4. Results of our calculations are shown 
in Fig. \ref{Tc_comp}. We can see that the experimentally observed \cite{UFN} 
high values of $T_c\sim$ 60--80K can be obtained only for large enough values of
the coupling constant of FeSe electrons with high -- energy optical phonon of 
STO $\lambda_2>$ 0.5, so that the total pairing coupling constant  
$\lambda=\lambda_1+\lambda_2>$ 0.9. Strictly speaking, such values of the
coupling constants can not be considered something unusual. However, the 
appearance of these large values in  FeSe/STO system seems rather improbable
in the light of qualitative estimates of $\lambda$ for nonadiabatic case in
Ref. \cite{Sad_18}, as well as the results of {\sl ab initio} calculations of
$\lambda$ for this system \cite{Johnson}. Note also, that the values of the
parameters used here for FeSe/STO belong to the intermediate region between
adiabatic or nonadiabatic regions, where our expressions, as was stressed above,
are of interpolating nature. Variation of the values of these parameters in
relatively wide range does not lead to the qualitative change of our results.
Traditionally low values of $\mu^{\star}$ used here, can not be obtained for
the assumed values of $D=E_F$, $\omega_1$ and coupling constants from expressions
like (\ref{mstar}) with usual values of $\mu$, due to rather small values of
corresponding Tolmachev's logarithm.

The author is grateful to E.Z. Kuchinskii for discussions and help with numerical
calculations. This work was partially supported by RFBR grant No. 17-02-00015 and
the program of fundamental research No. 12 of the RAS Presidium ``Fundamental
problems of high -- temperature superconductivity''.

\end{document}